\documentclass[aps,prl,twocolumn,showpacs,floatfix]{revtex4-1}
\usepackage{amsmath}

\usepackage{graphicx}

\begin{document}

\title{Time-periodic inertial range dynamics}

\author{Lennaert van Veen}
\email{lennaert.vanveen@uoit.ca}
\affiliation{Faculty of Science, University of Ontario Institute of Technology, 2000 Simcoe St. N., Oshawa, L1H 7K4
Ontario, Canada}

\author{Alberto Vela-Mart\'in}
\email{alberto@torroja.dmt.upm.es}
\affiliation{School of Aeronautics, Universidad Polit\'ecnica de Madrid, 28040 Madrid, Spain}

\author{Genta Kawahara}
\email{kawahara@me.es.osaka-u.ac.jp}
\affiliation{
Graduate School of Engineering Science,
Osaka University, 1-3 Machikaneyama, Toyonaka, Osaka 560-8531, Japan}

\date{\today}

\begin{abstract}
We present an unstable periodic orbit in large eddy simulation of an incompressible fluid in a periodic box
subject to a constant body force. The width of the inertial range of spatial scales, on which this simulation models high-Reynolds-number 
turbulence, is about three quarters of a decade, and a significant $-5/3$ scaling range is observed. We identify events of intense energy transfer across spatial scales and relate them to vortical dynamics. 
\end{abstract}


\maketitle

\noindent{\bf Introduction}\hspace{10pt} One of the defining properties of turbulence is the distribution of kinetic energy over a range of spatial scales according to a power law. Kolmogorov \cite{Kolmo} derived this power law using dimensional analysis under the assumptions that, on scales smaller than that on which the fluid is forced, e.g. by interaction with material boundaries or by a body force, but larger than that on which viscous damping is dominant, the flow is statistically isotropic, homogeneous and determined entirely by the rate of energy dissipation. The range of scales on which these assumptions hold approximately is called the inertial range, and the corresponding energy spectrum is called the ``-5/3 spectrum''. While scaling laws for the energy spectrum, velocity correlation functions and similar quantities
predict various time or ensemble averaged properties of turbulence, they give preciously little information about the fluid dynamics. A question that has become central to turbulence research since the publication of Kolmogorov's seminal works is: what dynamical processes give rise to the -5/3 spectrum?\\
The beginning of an answer is that turbulence is not featureless
but populated by multiple structures, such as vortex tubes and sheets, that stay coherent over long enough times to consider them ``building blocks''.
Such structures are distributed across scales and their interaction under Navier-Stokes dynamics is thought to be essential in the processes of the energy cascade.
One picture consistent with a power law spectrum, often attributed to Richardson \cite{Richardson1922}, is that such coherent structures break down in a self-similar fashion, thereby transfering energy to progressively smaller scales. Various mechanisms have been proposed for such break-down, for instance the ejection of thin, spiral vortex filaments from a large-scale vortex tube \cite{Lundgren1982}, the generation of counter-rotating vortex pairs wrapped around such a structure \cite{Melander1993} and, more recently, the iterated flattening and roll-up of vortex filaments \cite{Brenner2016}. What these mechanisms have in common, is that the transport of a given quantity of energy from the largest scale of the fluid motion down to that on which dissipation dominates takes a finite amount of time. This time delay explains the quasi-cyclic behaviour of spatial mean quantities, like the total energy and its rate of dissipation, observed both in experiments and in simulations (e.g. \cite{Pinton1999,cardesa2015,Goto2017}). In the first phase of the cycle, large-scale space structures grow under the influence of the forcing. In the next phase, these structures break down, thus supporting the energy cascade. Finally, the flow enters a quiescent state with little structure. This process repeats with a period much greater than the large-eddy turnover time. In various contexts the question has been posed whether this regeneration cycle of turbulence 
could be represented by a time-periodic solution to the governing equations \cite{YGK,Cvitanovic2013,Brenner2016}. 
In this Letter, we answer this question to the affirmative by presenting an Unstable Periodic Orbit (UPO) in Large Eddy Simulation (LES) of box turbulence, i.e. flow in a box with periodic boundary conditions.
To the best of our knowledge, this is the first such invariant solution ever computed in a flow with a significant inertial range of spatial scales.
We show the statistical properties of the time-periodic inertial range dynamics to be similar to those of turbulent Navier-Stokes flow, and
report the observaton of intermittent direct energy transfer events in physical space, energy backscatter and a time-delay between the energy injection and the energy dissipation signals. 

{\bf LES of box turbulence}\hspace{10pt} The Direct Numerical Simulation (DNS) of fully turbulent Navier-Stokes flow takes millions of degrees of freedom, which exceeds the limits of our current numerical methods for computing UPOs. We mitigate this problem by modelling the effect of the small-scale motion by an effective eddy viscosity. The resulting equations are
\begin{eqnarray}\label{LES}
\partial_t u+u\!\cdot\!\nabla u +\nabla \left(\frac{p}{\rho}+\frac{1}{3}\Pi\right) -2\nabla(\nu_T S)=\gamma f, \\
\nabla\!\cdot\! u=0,
\end{eqnarray}
where $u$, $p$ and $S$ are the grid-scale velocity, pressure and rate-of-strain tensor, respectively, and $\Pi$ contains the normal sub grid stress. The density, $\rho$, and the
amplitude of the force, $\gamma$, are constant. Using the closure
proposed by Smagorinsky \cite{Smagorinsky1963}, the eddy viscosity is
given by 
\begin{equation}\label{Smodel}
\nu_T=(C_{\rm S} \Delta)^2 \sqrt{2 S_{ij} S_{ij}}\,,
\end{equation}
where $C_{\rm S}$ is the 
Smagorinsky parameter, $\Delta$ is the grid spacing and summation over repeated indices is implied. While energy is removed from the system in an artificial way, the Smagorinsky closure has been shown to faithfully reproduce inertial range dynamics and statistics in the presence of periodic boundary conditions \cite{Lesieur1996,Linkmann2018}.\\ 
The force is $f=(-\sin(x)\cos(y),\cos(x)\sin(y),0)^T$, identical to that used by Yasuda {\sl et al.} \cite{YGK}, on a periodic domain of dimensions $2\pi\times 2\pi\times 2\pi$. The governing equations inherit the symmetries of this force so they are equivariant under translations in the vertical ($z$) direction.
Results below are scaled with the integral length scale,
\begin{equation}
L=\left\langle \frac{3\pi}{4 K}\int_0^{\infty} k^{-1} E(k,t)\,\mbox{d}k \right\rangle ,
\end{equation} 
the root-mean-square velocity, given by $U^2=2 \langle K\rangle/3$, and the large-eddy turnover time, $T=L/U$. Here, $K$ is the spatially averaged kinetic energy, $E(k,t)$ is the energy spectrum and $\langle . \rangle$ stands for the turbulent space-time average. The spatially averaged rate of energy transfer to sub grid scales is denoted by $\epsilon$.
We simulate this system with a pseudo-spectral code on a grid of $64^3$ points with $C_{\rm S}=0.55$. The correspnding ratio of the integral length scale to the LES filter length is $L/(C_{\rm S} \Delta)\approx 24$. 

\begin{figure}[t]
\includegraphics[width=.45\textwidth]{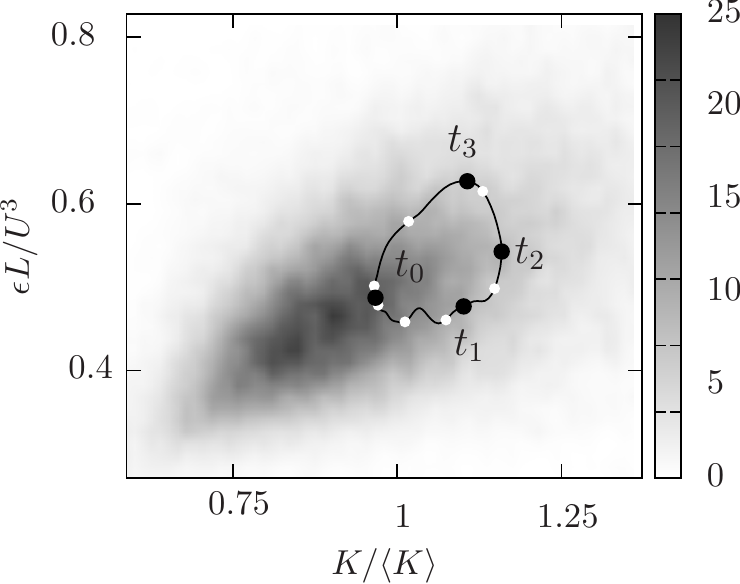}
\caption{Projection of the UPO onto $K$ and $\epsilon$, normalized by large scales. In grey scale the Probability Density Function (PDF) of turbulent LES is shown. The four time instants labeled $t_{0}$\ldots $t_{3}$, indicated by black dots, correspond to those in Fig. \ref{specs}(bottom). The white dots have been drawn at regular time intervals $9T/7$.}
\label{phase_space}
\end{figure}
\begin{figure}[t]
\includegraphics[width=.4\textwidth]{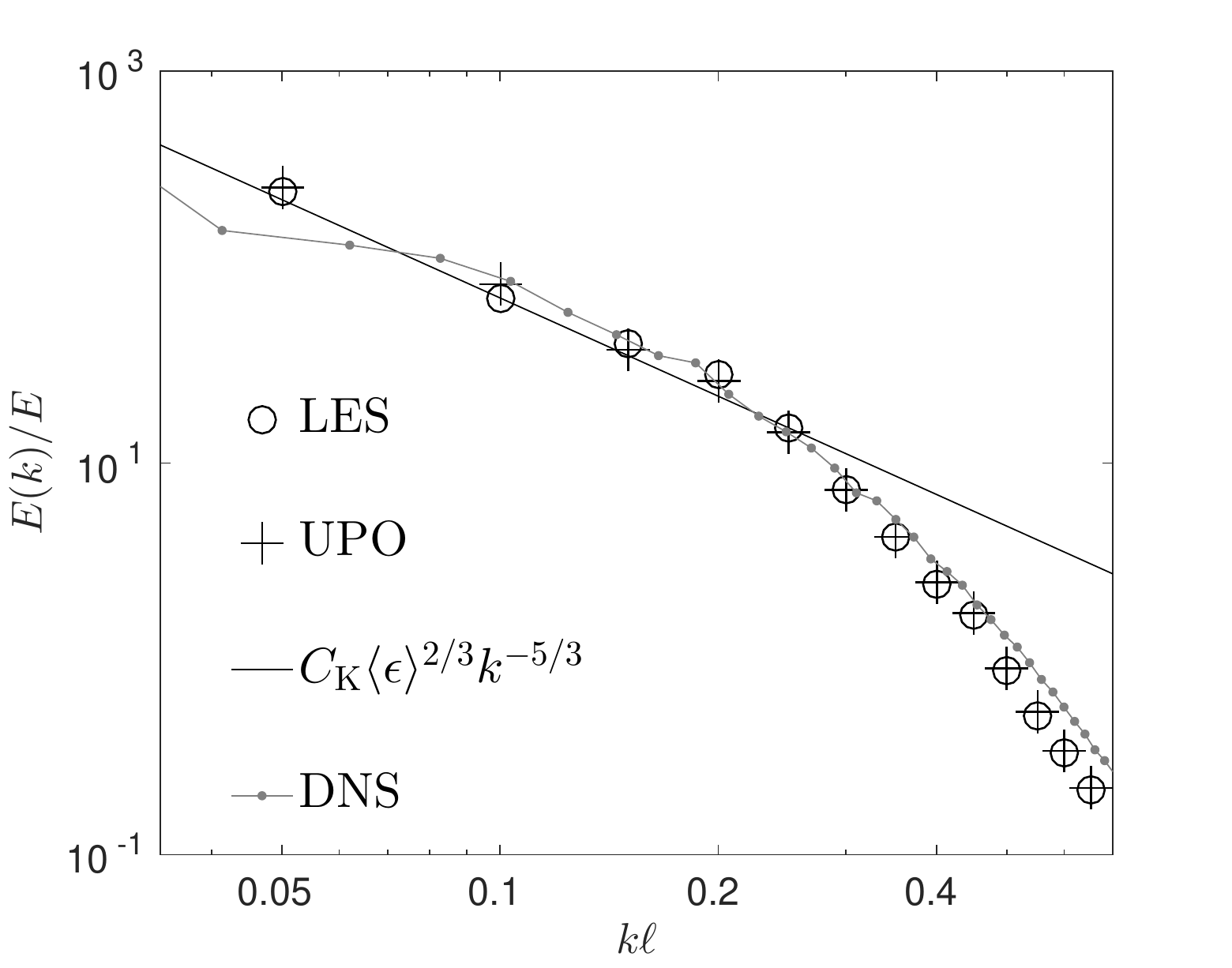}
\includegraphics[width=.4\textwidth]{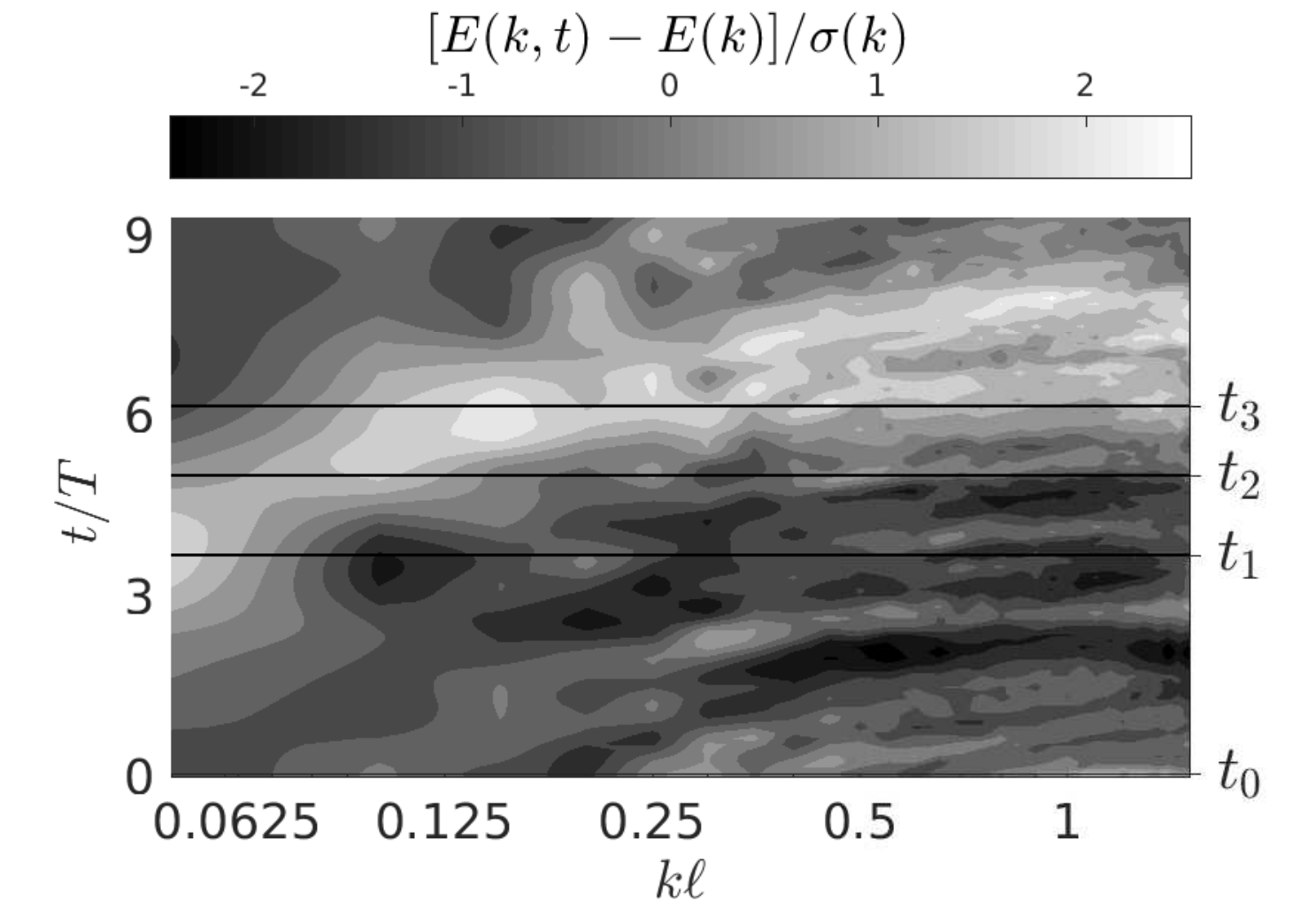}
\caption{The time-mean (top) and time-resolved (bottom) energy spectrum of the UPO. For comparison the time-mean spectrum of turbulent LES and DNS, the latter with $Re_{\lambda}=111$, have been included, as well as the theoretically expected Kolmogorov spectrum with $C_{\rm K}=1.5$. The time-resolved spectrum is visualized using the deviation of instantaneous spectrum $E(k,t)$ from the time-mean spectrum $E(k)$, normalized by the root-mean-square in time of $E(k,t)$ in LES turbulence, $\sigma(k)$. Here $k$ is the magnitude of the wave vector. The small-scale nondimensionalization is given by $E=[\langle \epsilon\rangle \langle \nu_{T}\rangle^5]^{1/4}$ and $\ell=[\langle \nu_T\rangle^3/ \langle \epsilon\rangle]^{1/4}$ for LES and $E=[\langle D\rangle \nu^5]^{1/4}$ and $\ell=[\nu^3/ \langle D\rangle]^{1/4}$ for DNS, where $D$ is the energy dissipation rate.}
\label{specs}
\end{figure}
\begin{figure}[t]
\includegraphics[width=0.5\textwidth]{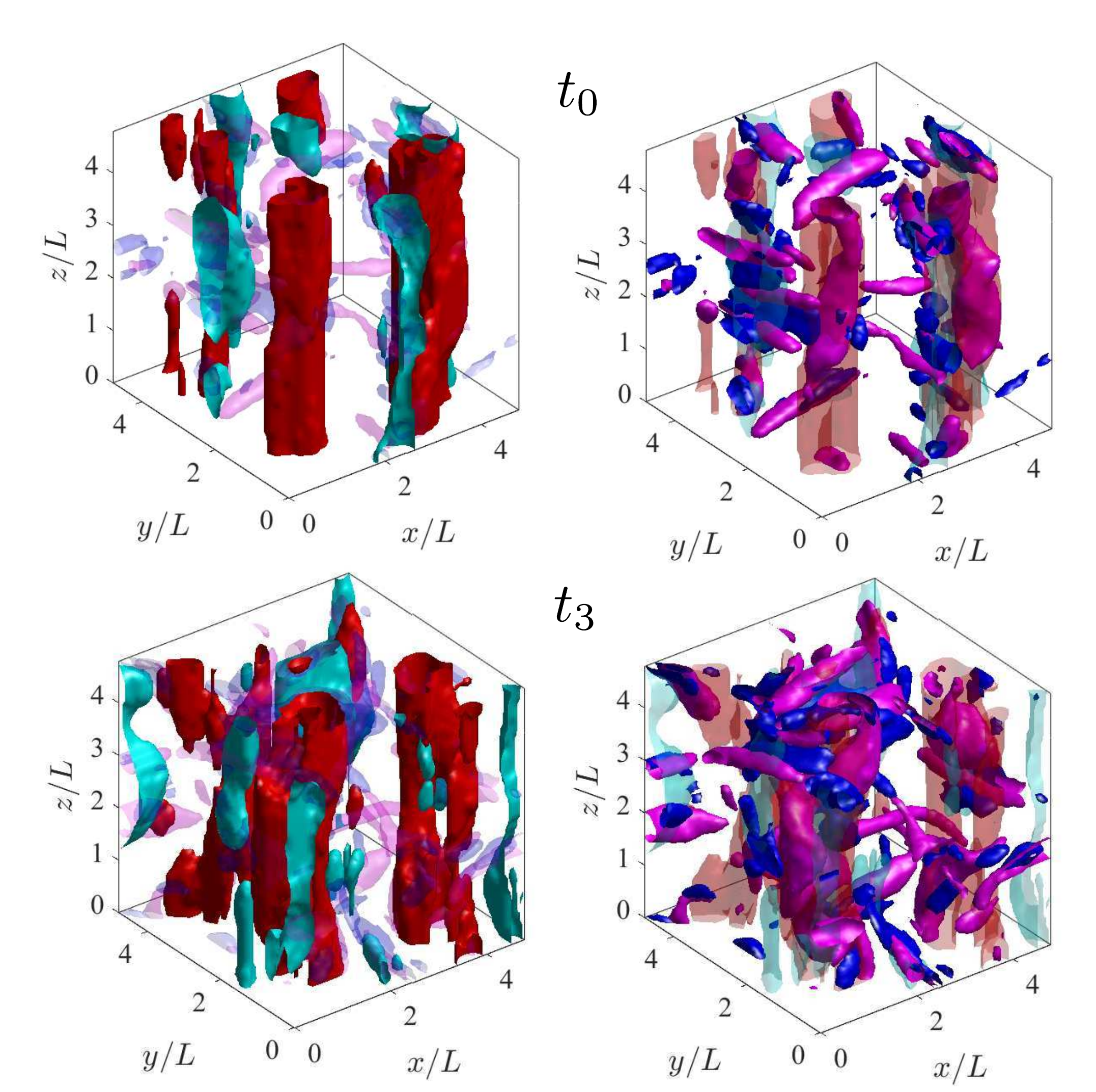}
\caption{Visualization of the spatial structure of the energy cascade at $t=t_0$ (top) and $t_3$ (bottom).
         (left) Highlighting the isosurface of intense enstrophy of the filtered field $\overline{\omega}_i^2 = 2.5\langle\overline{\omega}_i^2\rangle$ (red) and energy transfer $s(\boldsymbol x,t_3;0.8L)=3.0\langle s(0.8L) \rangle$ (cyan).
         (right) Highlighting the isosurface of enstrophy ${\omega}_i^2=2.5\langle {\omega}_i^2 \rangle$ (magenta) and 
         sub-grid energy transfer $s(\boldsymbol x, t_3; \Delta)=3.0\langle s(\Delta) \rangle$ (blue).} 
\label{cascade}
\end{figure}

\noindent{\bf Embedded time-periodic motion}\hspace{10pt} We employed a conventional Newton-Krylov-hook algorithm \cite{viswanath07} to locate a UPO in the LES flow. Starting from an approximately time-periodic segment of simulation data, this algorithm iteratively minimizes the distance from the final to the initial state, modulo a shift in the vertical direction. The rate of convergence is linear, and it took several thousands of iterations for the residual to reach the final value of $1.8\times 10^{-4}$, as measuerd by the energy of the difference between the final and the shifted initial state, normalized by the amplitude of the fluctuation of the energy along the UPO.\\
The resulting UPO has a period of $9T$ and a shift in the vertical of $0.85\Delta$ in one period. It has $210\pm 2$ unstable Floquet multipliers, the uncertainty arising from the finite residual of the UPO. The largest multiplier in magnitude corresponds to a decorrelation time of $1.6 T$.
A projection of the UPO onto $K$ and $\epsilon$ is shown in Fig. \ref{phase_space}. The white dots, drawn at regular time intervals, show that the state of the UPO is close to the centroid of the PDF of turbulence for about one third of its period.
From $t=t_0$ to $t_1$, the large-scale vortex columns grow under the influence of the external force, while small-scale motion is damped by the eddy viscosity. From $t=t_1$ to $t_2$, the large-scale vortices continue to grow but also bend as smaller-scale vortices start to grow in strength. From $t=t_2$ to $t_3$, the rate of energy transfer grows to its maximum and it is in this phase that an intense cascase process takes place. From $t=t_3$ to $t_0$ the large scale vortices are weak and vortical structures on small scales are damped by the eddy viscosity. Snap shots of the spatial structure at $t=t_0$ and $t_3$ are shown in Fig. \ref{cascade}.\\
This cycle is also visible in the time-resolved energy spectrum shown in Fig. \ref{specs}(bottom). In this figure, cascade processes show up as white streaks, moving upward, i.e. forward in time, and to the right, i.e. towards smaller scales. The time-averaged spectrum has a significant inertial range, shown in Fig. \ref{specs}(top). Both the spectrum of the UPO and that of turbulent LES agree fairly well the spectrum of DNS of the same system, using finite molecular and zero large eddy viscosity,  at $Re_{\lambda}=111$, excluding the dissipation range in which LES and DNS are expected to differ. In these figures, we normalized by the small scales of Kolmogorov theory.

\noindent{\bf The time-periodic energy cascade}\hspace{10pt} 
The examination of the spatio-temporal structure of the UPO reveals complex dynamics similar to those of developed turbulence.
We corroborate the presence of an energy cascade by analyzing the dynamics of the flow at different scales.


We filter the velocity field and calculate energy fluxes in physical space,
\begin{equation}
s(\boldsymbol x,t;l)= \tau_{ij} \overline{S}_{ij},
\end{equation}
where $\boldsymbol x$ is the spatial coordinate of the flow,
 $\overline{\cdot}$ denotes filtering at scale $l$ with a Gaussian filter in Fourier space, $G(k)=\exp(-k^2 l^2/24)$ \citep{aoyama2005statistics},
and $\tau_{ij}= \overline{u}_i\overline{u}_j - \overline{u_i u_j}$ are the sub-grid stresses at scale $l$.
Negative $s(\boldsymbol x, t)$ denotes energy flowing towards the small scales. At scale $\Delta$, the energy flux is
$s(\boldsymbol x,t;\Delta)=-2\nu_T S_{ij}S_{ij}$ and we have that $\langle \epsilon\rangle=-\langle s(\boldsymbol x,t;\Delta)\rangle$. 
The volume average of $s(\boldsymbol x,t;l)$ is denoted by $\Sigma(t;l)$. Also the enstrophy  $\omega_i^2$ and the enstrophy of the filtered field $\overline{\omega}_i^2$ are considered in this analysis.

We identify at least two hierachies of vortices, visualized at $t=t_0$ and $t_3$ in Fig. \ref{cascade}.
The first is generated by the 
forcing, and takes the form of four counter-rotating vortex columns.
Although constrained by the forcing, these column vortices change considerably in intensity, wind and meander, interact with the
next generation of vortices and display complex, three-dimensional, dynamics in time. 
The next generation of vortices populate the vicinity of the first generation and appear mostly perpendicular to these.
This interstitial region of the flow between the columnar vortices is characterized by a strong magnitude of the rate-of-strain tensor, 
which tends to develop a quasi-2D structure \cite{jimenez1992kinematic} and
stretch the second generation of vortices
perpendicular to the large columnar vortices.
Intense energy transfer at the scale of the large vortices
is also located in this region,
which suggests a connection of this mechanism with the energy cascade.
We observe that the small scale vortices reproduce the same dynamics of the first generation, creating strong strain in their vicinity, where the sub-grid model acts to remove energy from the resolved scales.
Intense sub-grid energy transfer
events are found to lie between vortical structures of the second generation. We would expect this mechanism to
repeat in a self-similar fashion in the presence of a wider inertial range.

\begin{figure}[t]
\includegraphics[width=0.4\textwidth]{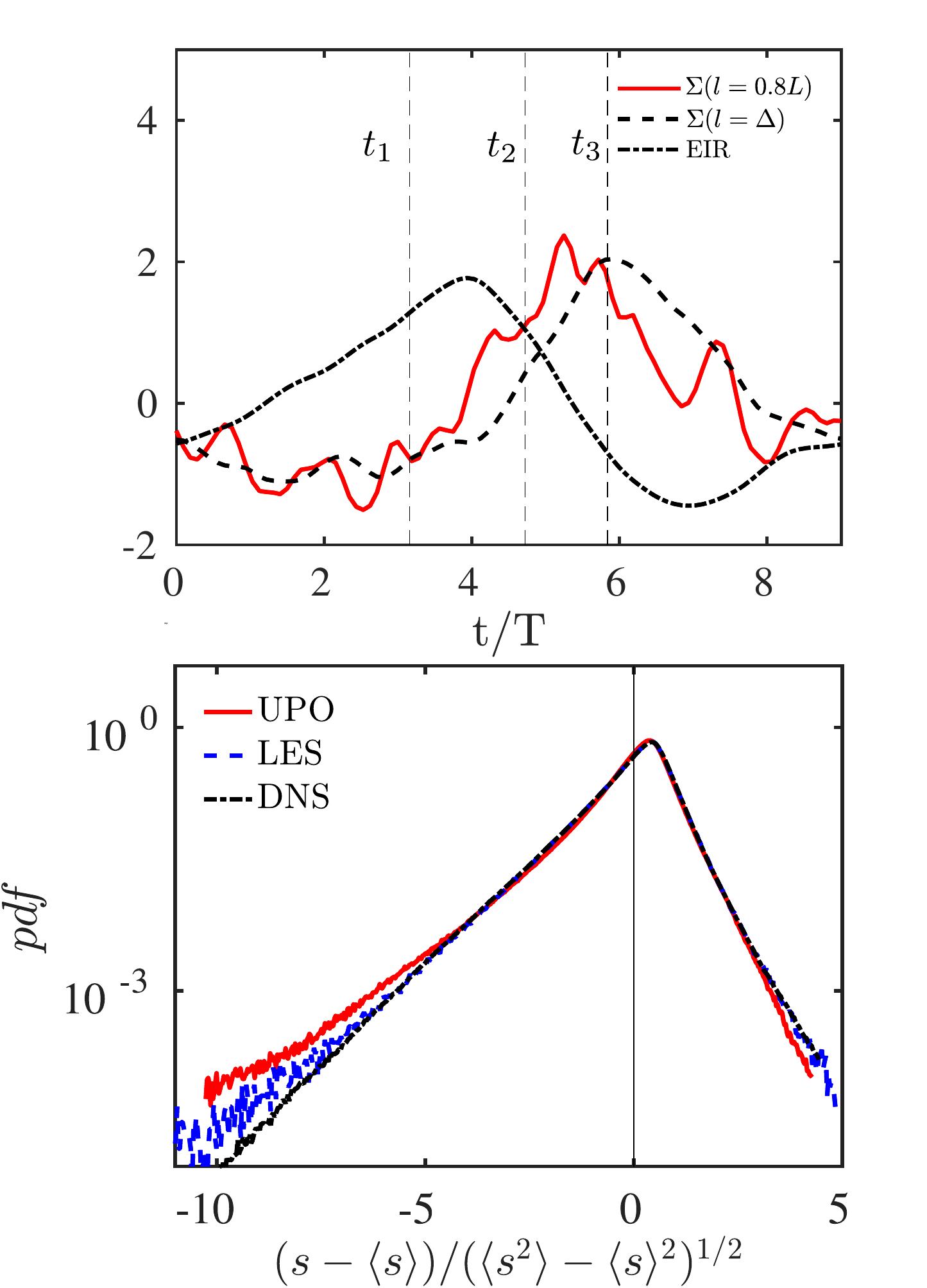}
\caption{(top) Time evolution of the volume-averaged energy injection rate (EIR), 
         the volume-averaged energy transfer $\Sigma(t)$ at scale $l=0.8L$ and $l=\Delta$, starting from $t_0=0$. 
         Signals are normalized by subtracting the temporal mean and dividing by the standard deviation in time.
         (bottom) Probability density function of energy transfer events $s(\boldsymbol x,t;0.8L)$ for the UPO, the LES flow and DNS flow at $Re_\lambda=111$.} 
\label{et_time}
\end{figure}

We find further evidences of the presence of at least a full step of the energy cascade in the analsis of the temporal evolution of the energy injection rate, the energy fluxes and the dissipation.
In Fig. \ref{et_time}(top), we show that the peak in  the energy injection rate precedes the peak in sub-grid energy transfer, while between we observe a peak in energy transfer.
This time-delay reproduces the cyclic dynamics observed in fully developed turbulence \cite{cardesa2015}, corroborating our analysis on the temporal evolution of the energy spectrum and evidencing the
presence of an inertial gap between the energy-injection scales and the small scales.
Intermittency, a key feature of the turbulence cascade,
is reproduced in the statistical distribution of $s(\boldsymbol x,t; 0.8L)$, which show wide tails towards negative values in Fig. \ref{et_time}(bottom). Energy backscatter, which is observed in fully developed turbulence \cite{aoyama2005statistics}, also suggests the presence of healthy inertial-range dynamcs.
As was the case for the energy spectrum, Fig. \ref{specs}(top), the data for the UPO, turbulent LES and DNS collapse, at least within five standard deviations from the mean, providing further evidence that the time-periodic intertial range dynamics faithfully represent that of turbulence. 

\noindent{\bf Conclusion}\hspace{10pt} We have presented a UPO in turbulent LES and have demonstrated that, although the intertial range is narrow, the periodic dynamics bear the hallmarks of the energy cascade process: a time delay between the maxima of energy input, transfer at intermediate scales and dissipation; spatial intermittency and the $-5/3$ scaling in the energy spectrum. The vortical dynamics observed in the UPO, and their overlap with energy transfer events, are consistent with the scenario proposed by Melander \& Hussain \cite{Melander1993} and investigated in detail by Goto {\sl et al.} \cite{Goto2017}.\\
We are not aware of other results on UPOs in LES, other than the prequel to the current work, \cite{VVKY}, and the study of plane Couette flow by Sasaki {\sl et al.} \cite{Sasaki2016}. The latter reproduced results obtained earlier in DNS, but did not reach a high enough separation of scales to observe the scaling laws typical of wall-bounded turbulence in the UPO. Sekimoto and Jim\'enez \cite{Sekimoto2017} studied a system with periodic boundary conditions, like the box turbulence studied here, but with constant shear imposed rather than an external force. At a scale separation comparable to ours they computed several travelling wave solutions, but these do not exhibit Kolmogorov scaling, possibly because their dynamics are too restricted.\\
The results reported here required several months of computing on modern GPU cards, due to the poor conditioning of the linear problems associated with Newton's method and the slow convergence of Krylov subspace iteration. We expect that significant improvements to the algorithms will be needed in order to compute a large number of UPOs in turbulent systems or even a single UPO with an inertial range spanning a full decade. 
The purpose of such computations is to gather detailed information about the dynamics of the energy cascade. In this approach, based on Periodic Orbit Theory (POT) \cite{Cvitanovic2013}, the UPOs are thought of as ``templates'' of turbulence. While they are much more expensive to compute than coherent structures extracted, for example, from Proper Orthogonal Decomposition, they convey proportionally more information since they are dynamical solutions to the governing equations. We hope that the current work will serve as a proof of principle that such templates of turbulence can indeed by found, as well as the starting point of an investigation of cascade dynamics based on tracking vortical structures, Lyapunov vectors and other quanities that can readily be computed for UPOs.

\noindent{\bf Acknowledgements}\hspace{10pt} This research was partially funded by the COTURB program of the European Research Council (ERC-2014.AdG-669505). LvV was supported by an NSERC Discovery Grant (nr. 355849-2013). GK was supported by the Grant-in-Aid for Scientific Research program of the Japan Society for the Promotion of Science (nos. 25249014, 26630055).
The authors gratefully acknowledge the computer resources at Minotauro and the technical support provided by Barcelona Supercomputing Center (FI-2017-3-0034).

\bibliographystyle{apsrev4-1}
\bibliography{TPIRD_v3}

\end{document}